\documentclass[twocolumn]{aastex63}
\shorttitle{Efficiently updating Karhunen-Lo\`eve eigenimages}
\shortauthors{Long et al.}
\graphicspath{{./}{./}}

\newcommand{\ZKL}{\ensuremath{\mathbf{Z}^\mathrm{KL}}}
\newcommand{\diag}[1]{\ensuremath{\mathrm{diag}\left(#1\right)}}

\begin{document}

\title{Unlocking starlight subtraction in full data rate exoplanet imaging by efficiently updating Karhunen-Lo\`eve eigenimages}

\correspondingauthor{Joseph D. Long}
\email{josephlong@email.arizona.edu}

\author[0000-0003-1905-9443]{Joseph D. Long}
\affiliation{Steward Observatory, The University of Arizona, Tucson, Arizona, USA}

\author{Jared R. Males}
\affiliation{Steward Observatory, The University of Arizona, Tucson, Arizona, USA}



\begin{abstract}
Starlight subtraction algorithms based on the method of Karhunen-Lo\`eve eigenimages have proved invaluable to exoplanet direct imaging.
However, they scale poorly in runtime when paired with differential imaging techniques.
In such observations, reference frames and frames to be starlight-subtracted are drawn from the same set of data, requiring a new subset of references (and eigenimages) for each frame processed to avoid self-subtraction of the signal of interest.
The data rates of extreme adaptive optics instruments are such that the only way to make this computationally feasible has been to downsample the data.
We develop a technique that updates a pre-computed singular value decomposition of the full dataset to remove frames (i.e. a ``downdate'') without a full recomputation, yielding the modified eigenimages.
This not only enables analysis of much larger data volumes in the same amount of time, but also exhibits near-linear scaling in runtime as the number of observations increases.
We apply this technique to archival data and investigate its scaling behavior for very large numbers of frames $N$.
The resulting algorithm provides speed improvements of $2.6\times$ (for 200 eigenimages at $N = 300$) to $140 \times$ (at $N = 10^4$) with the advantage only increasing as $N$ grows.
This algorithm has allowed us to substantially accelerate KLIP even for modest $N$, and will let us quickly explore how KLIP parameters affect exoplanet characterization in large $N$ datasets.
\end{abstract}



\section{Introduction} \label{sec:intro}

Direct imaging of exoplanets and disks depends on our ability to remove the effects of starlight on the signal of interest, whether before the focal plane with a coronagraph, or after with data post-processing techniques. The Locally Optimal Combination of Images (LOCI) technique of \cite{loci} achieved gains in contrast through post-processing by combining a reference library of images to approximate the frame under analysis. Karhunen-Lo\`eve image projection (KLIP) and principal component analysis (PCA) techniques for starlight subtraction were introduced to the field of high-contrast imaging in 2012 by \citeauthor{soummer_klip} and \citeauthor{amaraquanz}, bringing improved performance relative to LOCI. The former described a scheme for computing a Karhunen-Lo\`eve (KL) basis for PSF modeling from reference image covariance matrices, and the latter the principal components of the references via the singular value decomposition (SVD). (These operations are equivalent, as shown in Section~\ref{sec:comparesvd}.)

To decorrelate the target signal from the star signal, differential imaging techniques are used. These exploit differing behavior of the two signals with angle in angular differential imaging (ADI, \citealt{marois_adi}), wavelength in spectral differential imaging (SDI, \citealt{biller_sdi}, \citealt{sparks_ssdi}), or polarization in polarization differential imaging (PDI). When combined with KLIP, the recorded realizations of the PSF are both the data to be starlight-subtracted and the reference library from which to construct a model. By computing a new KL basis for each frame while drawing references from the \emph{other} frames, subtraction of any signal of interest is minimized.

Unfortunately, this recomputation of the KL basis at each frame increases the computation time faster than quadratically with the number of frames used (Section~\ref{sec:theory}). As data volumes from extreme adaptive optics (ExAO) instruments increase, making the same analyses possible in comparable amounts of time will either require scaling up computing resources massively \citep[e.g.][]{findr}, or identifying algorithmic improvements.

The usual solution is to combine frames in a sequence until the computation time is acceptably short. In the case of ADI, image combination will necessarily either change the amount of sky rotation in a single frame or the statistical properties of the images, presenting challenges. Furthermore, \cite{macintosh2005} report most atmospheric speckles have a lifetime of less than one second in simulation. Short integrations capture information about short-lifetime speckles, but they will be combined in the coadded frames, potentially causing the KL basis to capture spurious correlations.

Fortunately, low-rank representations of many-dimensional data are broadly useful outside astronomy, and their efficient computation is the subject of much work in machine learning and image processing. For instance, ``online'' algorithms like incremental PCA \citep{ross2008} enable incremental computation of outputs as data arrives without ever storing the entire data set in memory at once. Within astronomy, \cite{savransky_sequential_2015} explored the sequential calculation of covariance matrices and their inverses used in LOCI. In this work we apply a different mathematical identity to the SVD with the same goal of enabling analyses of larger datasets with only a linear increase in runtime.

Low-rank SVD modifications have applications not only to data postprocessing but also to control of the adaptive optics system that acquires the data. At the core of the empirical orthogonal functions formalism for predictive wavefront control described in \cite{guyoneof} lies an SVD that is iteratively updated to include new information on changing observatory conditions, for example.

Removing data from the low rank representation after the fact, however, has not been explored in as much detail (\citealt{hall2000}, \citealt{brand2006}). By describing the Karhunen-Lo\`eve transform in terms of an SVD of the image data, we develop an algorithm that uses the SVD of the full dataset to efficiently obtain the truncated decomposition of that dataset modified by the addition or removal of observations. This enables $\sim 2-100\times$ speedups that grow with the size of the data set (when compared with existing algorithms), enabling fast analyses of the full data volume from an ExAO system without compromises.

\section{The method of Karhunen-Lo\`eve image projections (KLIP)} \label{sec:background}

To define our terms and describe KLIP in terms of matrix operations, we restate the core algorithm of \cite{soummer_klip}. We begin with a target image $T$ and $N$ reference images $\{R_1, ..., R_N\}$ that have had a mean image subtracted (such that the mean value of each pixel taken separately through the stack of images is zero). Each observation is rearranged into a $\mathbb{R}^p$ vector of pixel values through an arbitrary mapping of $p$ pixel locations to vector elements, giving us the column vector $\vec{t}$ for our target image $T$ and a matrix $\mathbf{R}$ whose columns are $\left[\vec{r}_1, ..., \vec{r}_N\right]$ for our reference images.

For our reference set $\mathbf{R}$, the discrete Karhunen-Lo\`eve (KL) transform takes a series of $p$ dimensional vectors that represent observations of $p$ variables and transforms them into an orthogonal $p$-dimensional basis such that truncating the basis at the first $k$ vectors minimizes the mean squared error in reconstruction of the original data.
To subtract starlight without subtracting planet light, we exclude our target vector $\vec{t}$ (and, optionally, frames adjacent in angle or wavelength) from the reference set and we truncate the KL basis at $k$ vectors, or modes.

To obtain the KL transform, we first compute the image-to-image covariance matrix $\mathbf{C} = \mathbf{R}^T \mathbf{R}$. We then solve for its eigenvectors $\mathbf{V} = [\vec{v}_1, ..., \vec{v}_N]$ and corresponding eigenvalues $e_1 > ... > e_N$ such that $\mathbf{V}^T \mathbf{C} \mathbf{V} = \diag{\vec{e}} = \diag{e_1, ..., e_N}$.

Let $\diag{\vec{e'}} = \diag{e_1^{-1/2}, ..., e_N^{-1/2}}$. The optimal KL basis is then obtained by

\begin{equation}\mathbf{Z}^\mathrm{KL} = \mathbf{R} \mathbf{V} \diag{\vec{e'}}.\label{eq:klbasis}\end{equation}

The columns of \ZKL, when rearranged into images, are called the eigenimages of $\mathbf{R}$. To subtract starlight, we first define $\ZKL_k$ as the first $k$ columns of \ZKL, truncating the basis set to its $k$ most significant components. The estimate of the starlight in $\vec{t}$, $\tilde{t}$, is then

\begin{equation}\tilde{t} = \ZKL_k (\ZKL_k)^T \vec{t}\label{eq:starlight-estimate}\end{equation}

and the residual signal $\vec{a}$ is

\begin{equation}\vec{a} = \vec{t} - \tilde{t}.\label{eq:starlight-sub}\end{equation}

By mapping the entries of $\vec{a}$ back to pixel indices, the final image is obtained. These can then be combined, e.g. by median or rotate-and-stack ADI, to further enhance the signal-to-noise ratio of any astrophysical sources of interest.

\section{Equivalence of image-to-image covariance and direct SVD} \label{sec:comparesvd}

It is often more computationally efficient to compute \ZKL{} by finding the eigenvectors of the image-to-image covariance matrix (see Section~\ref{sec:theory}). However, the SVD is mathematically equivalent. Suppose $\mathbf{A} \in \mathbb{R}^{m \times n}$. Its SVD is then given by the orthogonal matrices $\mathbf{U} \in \mathbb{R}^{m \times m}$ and $\mathbf{V} \in \mathbb{R}^{n \times n}$ such that
$\mathbf{U}^T\mathbf{A}\mathbf{V} = \mathbf{\Sigma} = \diag{\sigma_1, ..., \sigma_q}$ where $q = \min(m,n)$ and $\sigma_1 \ge \sigma_2 \ge ... \ge \sigma_q$. By this definition, $\mathbf{A} = \mathbf{V} \mathbf{\Sigma} \mathbf{U}^T$, $\mathbf{U}\mathbf{U}^T = \mathbf{I}$ and $\mathbf{V}\mathbf{V}^T = \mathbf{I}$, so

\begin{eqnarray*}
    \mathbf{A}^T \mathbf{A} =& (\mathbf{V} \mathbf{\Sigma}^T \mathbf{U}^T) (\mathbf{U} \mathbf{\Sigma} \mathbf{V}^T)\\
    \mathbf{A}^T \mathbf{A} =& \mathbf{V} \mathbf{\Sigma}^T \mathbf{\Sigma} \mathbf{V}^T\\
    \mathbf{A}^T \mathbf{A} =& \mathbf{V} \diag{\sigma_1^2, ..., \sigma_q^2} \mathbf{V}^T\\
\end{eqnarray*}

If we set $\mathbf{A} = \mathbf{R}$, we see that this $\mathbf{V}$ and the $\mathbf{V}$ above that diagonalizes the covariance matrix $\mathbf{C}$ must be equivalent and that the singular values are related to the eigenvalues of the covariance by $\sigma_i^2 = e_i$. 
To compute \ZKL{} from this representation, we can write $\mathbf{\Sigma}^{-1} = \diag{(\sigma_1^2)^{-1/2}, ..., (\sigma_N^2)^{-1/2}} = \diag{\sigma_1^{-1}, ..., \sigma_N^{-1}}$. Equation~\ref{eq:klbasis} becomes

$$\mathbf{Z}^\mathrm{KL} = \mathbf{R} \mathbf{V} \mathbf{\Sigma}^{-1}.$$

Replacing $\mathbf{R}$ with its SVD gives

\begin{eqnarray*}
    \mathbf{Z}^\mathrm{KL} =& (\mathbf{U} \mathbf{\Sigma} \mathbf{V}^T) \mathbf{V} \mathbf{\Sigma}^{-1}\\
        =& \mathbf{U} \mathbf{\Sigma} \mathbf{I} \mathbf{\Sigma}^{-1} \\
        =& \mathbf{U}.
\end{eqnarray*}

Thus, the left singular vectors $\mathbf{U}$ of the reference matrix $\mathbf{R}$ are the basis \ZKL{} we computed in Section~\ref{sec:background}. 

For $N$ reference images and $p$ pixels where $p \gg N$, it is advantageous to compute \ZKL{} from the $N \times N$ image-to-image covariance matrix rather than the $N \times p$ matrix of vectorized images. Eigenvectorization of the image-to-image covariance is an $O(N^3)$ problem, while the SVD requires $O(pN^2)$ operations. For data where $N$ exceeds $p$, the SVD is preferred as the matrix can always be transposed such that the cost is $O(Np^2)$.

The case where $N$ exceeds $p$ may arise not only when an exceptionally long sequence of images is taken, but also when the number of pixels $p$ is reduced through masking or dividing the image into search regions that each contain far fewer pixels than the entire image does.

\section{Modification of the SVD to obtain updated eigenimages} \label{sec:algorithm}

When processing differential imaging data from ground-based telescopes, $\vec{t}$ and $\mathbf{R}$ are just submatrices of a larger observations matrix $\mathbf{X} \in \mathbb{R}^{p \times (N + 1)}$ whose columns we will call $\mathbf{X} = [\vec{x}_1, ..., \vec{x}_{N+1}]$. While processing all $N + 1$ frames, each column of $\mathbf{X}$ will take its turn as the target $\vec{t}$ and all other columns will form the reference. Other exclusion schemes are possible, e.g. based on field rotation or wavelength difference to prevent overlap of an astrophysical signal. What follows removes of one column of $\mathbf{X}$, but will readily generalize to removal of multiple observations.

To process frame $i$, $$\vec{t} = \vec{x}_{i}\ \mathrm{and}\ \mathbf{R} = [\vec{x}_1, ..., \vec{x}_{i-1}, \vec{x}_{i+1}, ..., \vec{x}_{N+1}].$$ To compute $\tilde{t}$, we would find the SVD of $\mathbf{R} = \mathbf{U} \mathbf{\Sigma} \mathbf{V}^T$ and truncate to the first $k$ singular values and vectors (or, equivalently, compute a rank-$k$ thin SVD) to obtain $\mathbf{R} \approx \mathbf{U}_k \mathbf{\Sigma}_k \mathbf{V}_k^T$. Then, we would calculate Equation~\ref{eq:starlight-estimate} with $\ZKL_k = \mathbf{U}_{k}$.

We would repeat this procedure for every frame $i \in [1, N+1]$. This gives KLIP with differential imaging unfortunate $O(pN^3)$ scaling. (For calculation of \ZKL{} from the covariance, $O(N^4)$.) However, we can exploit the fact that the left singular vectors we recompute correspond to reference matrices $\mathbf{R}$ that differ from $\mathbf{X}$ only by the presence or absence of some columns.

\cite{brand2006} developed an identity for additive (and subtractive) modifications of an SVD, which we will apply to our matrices. Suppose that we have already computed the SVD of our full data set through a standard solver to obtain $\mathbf{X} = \mathbf{U}_0 \mathbf{\Sigma}_0 \mathbf{V}_0$. After truncation, this gives a low-rank approximation, $\mathbf{X} \approx \mathbf{U}_{0,k} \mathbf{\Sigma}_{0,k} \mathbf{V}_{0,k}^T$. We want to find the low-rank or truncated SVD of $\mathbf{R}$, that is, $\mathbf{X}$ without column $i$.

To modify the data matrix to remove $\vec{x}_i$, we express $\mathbf{R} = \mathbf{X} + \mathbf{A}\mathbf{B}^T$, the sum of $\mathbf{X}$ and the product of two low rank matrices $\mathbf{A} \in \mathbb{R}^{p \times c}$, $\mathbf{B} \in \mathbb{R}^{(N+1) \times c}$. The addition of $\mathbf{A}\mathbf{B}^T$ will zero $c$ columns of the resulting matrix by adding $-\vec{x}_i$ to column $i$ for every $i$ removed. It can be verified that the left singular vectors of an $m \times (n + 1)$ matrix with an excess column of zeros are identical to those for a $m \times n$ matrix where the column is omitted entirely---up to a sign factor. (This sign factor is unimportant for starlight subtraction, as the product $\ZKL_k (\ZKL_k)^T$ in Equation~\ref{eq:starlight-estimate} will cancel it out, but it should be accounted for when comparing singular vectors produced by different algorithms.)

In this case, $c = 1$, $\mathbf{A} = \vec{x}_i$, and $\mathbf{B}$ is a vector of zeros where the $i$th component has been set to $-1$. When removing multiple columns, each column of $\mathbf{A}$ corresponds to the contents of the column to remove, and each column of $\mathbf{B}$ has a $-1$ in the row corresponding to the index of the removed column in the original matrix.

The sum $\mathbf{X} + \mathbf{A}\mathbf{B}^T$ can be expressed in terms of $\mathbf{X}$'s SVD as
\begin{equation}
    \mathbf{X} + \mathbf{A}\mathbf{B}^T = [
        \mathbf{U}_{0}\ \mathbf{A}
        ] \left[
            \begin{array}{cc}
                \mathbf{\Sigma}_{0} & \mathbf{0}\\\mathbf{0} & \mathbf{I}
            \end{array}
        \right]
        [
            \mathbf{V}_{0}\ \mathbf{B}
        ]^T
    \label{eq:augmentsvd}.
\end{equation}

We want to modify the truncated SVD of rank $k$, so Equation~\ref{eq:augmentsvd} becomes

\begin{equation}
    \mathbf{X} + \mathbf{A}\mathbf{B}^T \approx [
        \mathbf{U}_{0,k}\ \mathbf{A}
        ] \left[
            \begin{array}{cc}
                \mathbf{\Sigma}_{0,k} & \mathbf{0}\\\mathbf{0} & \mathbf{I}
            \end{array}
        \right]
        [
            \mathbf{V}_{0,k}\ \mathbf{B}
        ]^T
    \label{eq:truncaugmentsvd}.
\end{equation}

The left and right block matrices may be rewritten as
\begin{eqnarray}
    [\mathbf{U}_{0,k}\ \mathbf{A}] =& [\mathbf{U}_{0,k}\ \mathbf{P}] \left[\begin{array}{cc}\mathbf{I} & \mathbf{U}_{0,k}^T\mathbf{A}\\\mathbf{0} & \mathbf{R_A}\end{array}\right]\label{eq:leftrightdecomp1}\\
    \ \mathrm{and}\nonumber\\
    \ [\mathbf{V}_{0,k}\ \mathbf{B}] =& [\mathbf{V}_{0,k}\ \mathbf{Q}] \left[\begin{array}{cc}\mathbf{I} & \mathbf{V}_{0,k}^T\mathbf{B}\\\mathbf{0} & \mathbf{R_B}\end{array}\right]\label{eq:leftrightdecomp2}
\end{eqnarray}

where $\mathbf{P}$ is an orthogonal basis of the column space of $(\mathbf{I} - \mathbf{U}_{0,k}\mathbf{U}_{0,k}^T)\mathbf{A}$ (which may be obtained by QR decomposition) and $\mathbf{R_A} = \mathbf{P}^T (\mathbf{I} - \mathbf{U}_{0,k}\mathbf{U}_{0,k}^T)\mathbf{A}$. $\mathbf{Q}$ and $\mathbf{R_B}$ are defined similarly for $\mathbf{V}_{0,k}$ and $\mathbf{B}$.

The additional column(s) in $\mathbf{P}$ capture the component(s) of $\mathbf{A}$ orthogonal to $\mathbf{U}_{0,k}$, and $\mathbf{R_A}$ represents that component in terms of $\mathbf{P}$. By substituting Equations~\ref{eq:leftrightdecomp1} \& \ref{eq:leftrightdecomp2} into Equation~\ref{eq:augmentsvd}, we obtain

\begin{equation}
    \mathbf{X} + \mathbf{A}\mathbf{B}^T \approx
        [\mathbf{U}_{0,k}\ \mathbf{P}]
        \mathbf{K}
        [\mathbf{V}_{0,k}\ \mathbf{Q}]^T
    \label{eq:k-in-decomp}
\end{equation}

where we have defined $\mathbf{K}$ as

\begin{equation}
    \mathbf{K} = \left[\begin{array}{cc}\mathbf{I} & \mathbf{U}_{0,k}^T\mathbf{A}\\\mathbf{0} & \mathbf{R_A}\end{array}\right]
        \left[\begin{array}{cc}\mathbf{\Sigma}_{0,k} & \mathbf{0}\\\mathbf{0} & \mathbf{I}\end{array}\right]
        \left[\begin{array}{cc}\mathbf{I} & \mathbf{V}_{0,k}^T\mathbf{B}\\\mathbf{0} & \mathbf{R_B}\end{array}\right]^T.
    \label{eq:initial-k}
\end{equation}

\cite{brand2006} notes the number of columns of $\mathbf{P}$ and number of rows of $\mathbf{R_A}$ are equal to the rank of this orthogonal component, and may be zero. Since we are removing columns of $\mathbf{X}$, our $\mathbf{A}$ matrix contains data already incorporated in $\mathbf{U}_{0,k}$ and our $\mathbf{B}$ matrix contains data already incorporated in $\mathbf{V}_{0,k}$. Evaluating $\mathbf{P}$, $\mathbf{R_A}$, $\mathbf{Q}$, and $\mathbf{R_B}$ when removing columns confirms that all their entries are very nearly zero. In fact, the component captured by $\mathbf{P}$ is exactly the residual starlight after subtracting the reconstruction, and incorporating it would produce an eigenbasis that almost perfectly subtracts $\vec{t}$. As we are interested in the residuals $\vec{a} = \vec{t} - \tilde{t}$, this is an undesirable effect. Therefore, we should ignore $\mathbf{P}$, $\mathbf{R_A}$, $\mathbf{Q}$, and $\mathbf{R_B}$, treating them as if they had zero rows and columns, simplifying Equations~\ref{eq:k-in-decomp}~\&~\ref{eq:initial-k} to:

\begin{equation}
    \mathbf{X} + \mathbf{A}\mathbf{B}^T \approx
        \mathbf{U}_{0,k}
        \mathbf{K}
        \mathbf{V}_{0,k}^T
    \label{eq:simple-k}
\end{equation}

and

\begin{eqnarray}
\mathbf{K} =& [\mathbf{I}\ \mathbf{U}_{0,k}^T\mathbf{A}] \left[\begin{array}{cc}\mathbf{\Sigma} & \mathbf{0}\\\mathbf{0} & \mathbf{I}\end{array}\right] \left[\begin{array}{c}\mathbf{I} \\ \mathbf{V}_{0,k}^T\mathbf{B}\end{array}\right]\nonumber\\
    =& [\mathbf{\Sigma}\ \mathbf{U}_{0,k}^T\mathbf{A}] \left[\begin{array}{c}\mathbf{I} \\ \mathbf{V}_{0,k}^T\mathbf{B}\end{array}\right]\nonumber\\
\mathbf{K} =& \mathbf{\Sigma} + \mathbf{U}_{0,k}^T\mathbf{A} \mathbf{V}_{0,k}^T\mathbf{B}.
\end{eqnarray}

To obtain the updated SVD $\mathbf{X} + \mathbf{A}\mathbf{B}^T = \mathbf{U}_k \mathbf{\Sigma}_k \mathbf{V}_k^T$ we must re-diagonalize $\mathbf{K}$ as $\mathbf{K} = \mathbf{W} \mathbf{\Sigma}_k \mathbf{Y}^T$ by computing an SVD of reduced cost $O(k^3)$.

By substituting this diagonalization for $\mathbf{K}$ into Equation~\ref{eq:simple-k}, we obtain

\begin{eqnarray}
    \mathbf{X} + \mathbf{A}\mathbf{B}^T = \mathbf{R} \approx&
        (\mathbf{U}_{0,k}
        \mathbf{W})
        \mathbf{\Sigma}_k
        (\mathbf{V}_{0,k} \mathbf{Y})^T
        \label{eq:svd-downdate}
        \\
      \approx& \mathbf{U}_k
        \mathbf{\Sigma}_k
        \mathbf{V}_k^T \nonumber.
\end{eqnarray}

By the definition of the SVD, $\mathbf{W}$ and $\mathbf{Y}$ are orthogonal matrices, and so are the products $\mathbf{U}_{0,k} \mathbf{W}$ and $\mathbf{V}_{0,k} \mathbf{Y}$. Therefore $\mathbf{U}_k = \mathbf{U}_{0,k} \mathbf{W}$ are the left singular vectors, $\mathbf{V}_k = \mathbf{V}_{0,k} \mathbf{Y}$ are the right singular vectors, and $\mathbf{\Sigma}_k$ is the diagonal matrix of the updated singular values.

Thus we have updated the SVD by removing data from, or ``downdating,'' the lower-rank decomposed representation of $\mathbf{X}$ directly. 


\section{Practical considerations for implementation} \label{sec:implementation}

\begin{deluxetable*}{CCl}
    \tablenum{1}
    \tablecaption{Summary of symbols and dimensions\label{tab:dimensions}}
    \tablewidth{0pt}
    \tablehead{
        \colhead{Symbol} & \colhead{Dimensions} & \colhead{Description}
    }
    \startdata
    N & & Number of frames used in the reference set\\
    p & & Number of pixels in a search region\\
    k & & Number of modes (eigenvectors or singular vectors) used in starlight subtraction\\
    \vec{t} & p & Mean-subtracted target image unwrapped into the entries of a vector\\
    \vec{r}_i & p & Mean-subtracted reference image unwrapped into the entries of a vector\\
    \mathbf{R} & p \times N & Matrix of mean-subtracted reference images\\
    \mathbf{C} & N \times N & Image to image covariance matrix\\
    \vec{e} & N & Vector of eigenvalues of $\mathbf{C}$ arranged in descending order\\
    \ZKL{} & p \times N & Karhunen-Lo\`eve basis for $\mathbf{R}$\\
    \ZKL_k & p \times k & Truncated Karhunen-Lo\`eve basis for $\mathbf{R}$\\
    \mathbf{U}_k & p \times k & First $k$ left singular vectors of $\mathbf{R}$, equal to $\ZKL_k$\\
    \mathbf{\Sigma}_k & k \times k & Diagonal matrix of the $k$ largest singular values of $\mathbf{R}$\\
    \mathbf{V}_k & N \times k & First $k$ right singular vectors of $\mathbf{R}$\\
    \vec{x}_i & p & A mean-subtracted image unwrapped into the entries of a vector\\
    \mathbf{X} & p \times N + 1 & Matrix of all mean-subtracted data\\
    c & & Number of columns to add or remove from $\mathbf{X}$\\
    \mathbf{A} & p \times c & Matrix whose columns are the columns $\vec{x_i}$ of $\mathbf{X}$ to be removed\\
    \mathbf{B} & N \times c & Matrix of zeros with a $-1$ in the row of each column that corresponds to the column index in $\mathbf{X}$ to be removed\\
    \mathbf{U}_{0,k} & p \times k & First $k$ left singular vectors of $\mathbf{X}$\\
    \mathbf{\Sigma}_{0,k} & k \times k & Diagonal matrix of the $k$ largest singular values of $\mathbf{X}$\\
    \mathbf{V}_{0,k} & N \times k & First $k$ right singular vectors of $\mathbf{X}$\\
    \mathbf{K} & k \times k & Matrix representing the original singular values and low-rank modification to be re-diagonalized\\
    \mathbf{W} & k \times k & Rotation of the left singular vectors multiplied on the left by $\mathbf{U}_{0,k}$ to obtain $\mathbf{U}_k$\\
    \mathbf{Y} & k \times k & Rotation of the right singular vectors multiplied on the left by $\mathbf{V}_{0,k}$ to obtain $\mathbf{V}_k$\\
    \enddata
\end{deluxetable*}

This algorithm is straightforward to implement with basic matrix operations and an SVD solver from a library (e.g. LAPACK, Intel MKL). Table~\ref{tab:dimensions} summarizes the symbols used above and their dimensions for the benefit of the implementer.

Superfluous operations can be eliminated by noting that the right singular vectors $\mathbf{V}_k$ are not necessary for KLIP, and the product $\mathbf{V}_{0,k}\mathbf{Y}$ need not be calculated. Furthermore, we have assumed that the rank-$k$ representation of the data captures all the information necessary for starlight subtraction, so the product $\mathbf{U}_{0,k}^T\mathbf{A}$ is just $-\mathbf{\Sigma}_{0,k}\vec{v}^T$ where $\vec{v}$ is the $i$th \emph{row} vector of $\mathbf{V}_{0,k}$. Similarly, $\mathbf{V}_{0,k}^T\mathbf{B}$ is just $\vec{v}^T$. (In effect, we are removing the vector $\mathbf{U}_{0,k} \mathbf{U}_{0,k}^T \vec{t}$ rather than $\vec{t}$.)

To prevent the erosion of numerical accuracy, it is helpful to compute the decomposition of $\mathbf{X} \approx \mathbf{U}_{0,k+1} \mathbf{\Sigma}_{0,k+1} \mathbf{V}_{0,k+1}$ and use $k + 1$ singular vectors and values through the algorithm in Section~\ref{sec:algorithm}, truncating to $\ZKL_k = \mathbf{U}_k$ just before evaluating Equation~\ref{eq:starlight-estimate}.  When removing more than one column per iteration, the rank of the initial decomposition should be increased by that number of singular values to mitigate errors at greater $k$.

We have omitted discussion of the difference in mean for each set of reference images, assuming the difference between the mean image of the entire set of references and the mean of a subset with some frames excluded is small. This approximation holds as long as the number of images removed, $c$, is much smaller than $N$.

Finally, note that the $\mathbf{U}_{k} \mathbf{\Sigma}_{k} \mathbf{V}_{k}$ computed in Equation~\ref{eq:svd-downdate} is the SVD not of a modification to $\mathbf{X}$ but rather to $\mathbf{U}_{0,k} \mathbf{\Sigma}_{0,k} \mathbf{V}_{0,k}^T$, and the two are not strictly equal unless $\mathrm{rank}(\mathbf{X}) < k$. The numerical experiments in Section~\ref{sec:performance} show that this approximation has little impact on the recovered signal-to-noise ratio of a planet.

\section{Performance} \label{sec:performance}

\subsection{Impact on recovered signal to noise} \label{sec:compare_snr}

\begin{figure}[ht!]
    \plotone{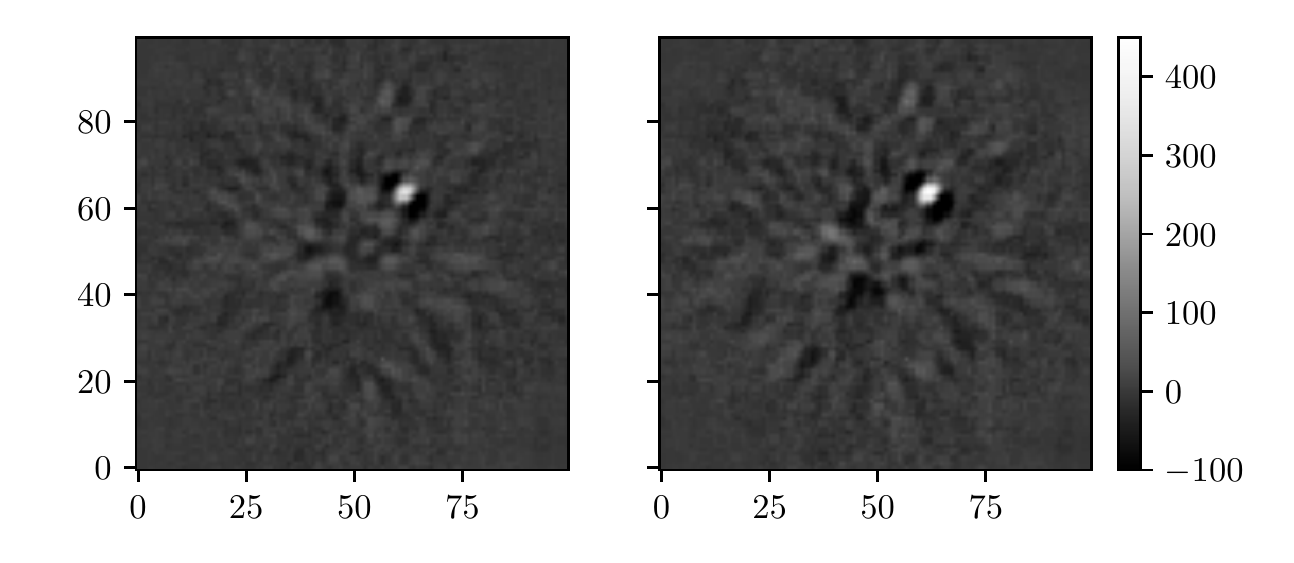}
    \caption{NACO ADI observations of $\beta$ Pictoris b processed by recomputing eigenimages for each frame (left) and by modifying the SVD (right), subtracting starlight, and combining the images by summing after derotation. In both cases, $k = 30$ modes, $N = 60$ reference frames (from a total of 61), $p = 10^4$ pixels. The structure in the residuals is not identical, and the SVD modification provides slightly better throughput, though this does not appear to be significant.\label{fig:sidebyside}}
\end{figure}

\begin{figure}[ht!]
    \plotone{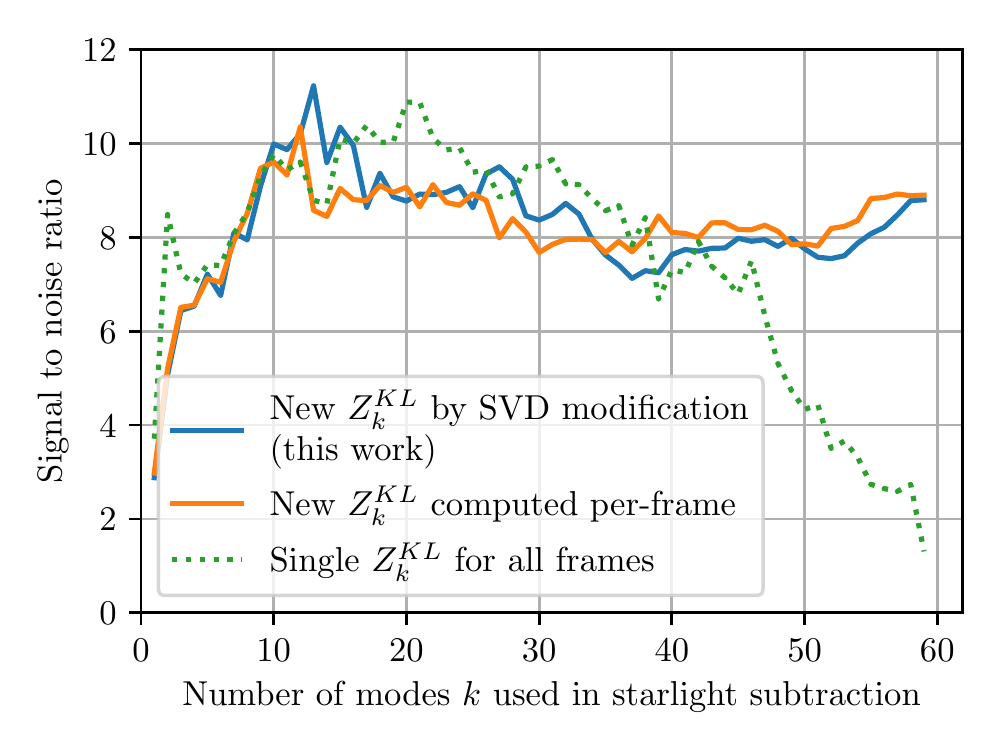}
    \caption{Comparison of recovered SNR for the data from Figure~\ref{fig:sidebyside} as a function of $k$ modes for KLIP with a single set of eigenimages computed from the entire dataset (without any frame exclusion), KLIP with $\ZKL_k$ recomputed at each frame from $\mathbf{C}$ with the target frame excluded, and KLIP with $\ZKL_k$ computed via SVD modification to exclude the target frame. In terms of recovered signal to noise ratio, the SVD modification algorithm from this work provides the same advantages as recomputing $\ZKL_k$ at each frame when compared to a naive KLIP with a single $\ZKL_k$ for all images.\label{fig:snr}}
\end{figure}


To evaluate the impact of algorithm choice on signal-to-noise ratio (SNR) with realistic data, we used observations of $\beta$ Pictoris b from \cite{nacodata} made available as part of the VIP software package \citep{VIP}. An example of the final KLIP+ADI output for the existing and modified SVD algorithms can be seen in Figure~\ref{fig:sidebyside}. The effect of algorithm choice and number of modes $k$ used is summarized by Figure~\ref{fig:snr}. The SVD modification algorithm does not result in numerically identical values for SNR at various $k$ modes when compared with direct computation of the singular vectors or eigenvectors, but we are satisfied that it performs comparably.

\subsection{Theoretical} \label{sec:theory}

As we are most interested in the case where number of observations exceeds number of pixels, we will assume $N > p$ in the following.

The cost of the image-to-image covariance calculation is $O(N^3 + pN^2)$ per frame, where the cubed term is due to the eigenvectorization \citep[][\S 8.3]{matrixcomputations} and the other term is due to the matrix product to calculate the covariance matrix $\mathbf{C}$. The pixel-to-pixel covariance calculation, useful when $N > p$, has complexity $O(p^3 + Np^2)$ per frame. When only $k$ eigenvectors are needed, the costs are $O(N^2k + pN^2)$ and $O(p^2k + Np^2)$, respectively.

The current state of the art in calculating intractably large SVDs is a randomized algorithm \citep{halko_finding_2011}, which can provide a highly accurate approximation of the SVD in $O(Np\log(k) + (N+p)k^2)$ time. Halko gives the scaling behavior for the conventional low-rank SVD as $O(Npk)$.

So far, we have discussed the computational cost of computing $\ZKL_k$ for each frame in isolation. When analyzing a sequence of frames, each of these estimates will be multiplied by an additional factor of $N$ for the number of frames on which the procedure is repeated to obtain the total cost. The algorithm we have developed sidesteps this additional $N$ factor by amortizing the cost of the initial decomposition over the $N$ frames.

The SVD modification algorithm we have applied to KLIP costs $O((N+1)p^2)$ for an initial deterministic SVD and $O(k^3 + pk^2)$ for each iteration of the re-decomposition and computation of $\mathbf{U}_k$ from $\mathbf{U}_{0,k}\mathbf{W}$. Dividing the full initial decomposition cost by the number of frames gives a complexity of $O(p^2 + k^3 + pk^2)$ for finding the eigenimages of a single frame. In other words, the total cost will still increase linearly with $N$, but the per-frame cost will remain constant. This is in contrast to the other algorithms mentioned above, in which the cost of processing a \emph{single} frame depends on the \emph{total} number of frames, with execution time growing quadratically (or faster) with $N$.

\subsection{Empirical}

\begin{figure*}
    \plotone{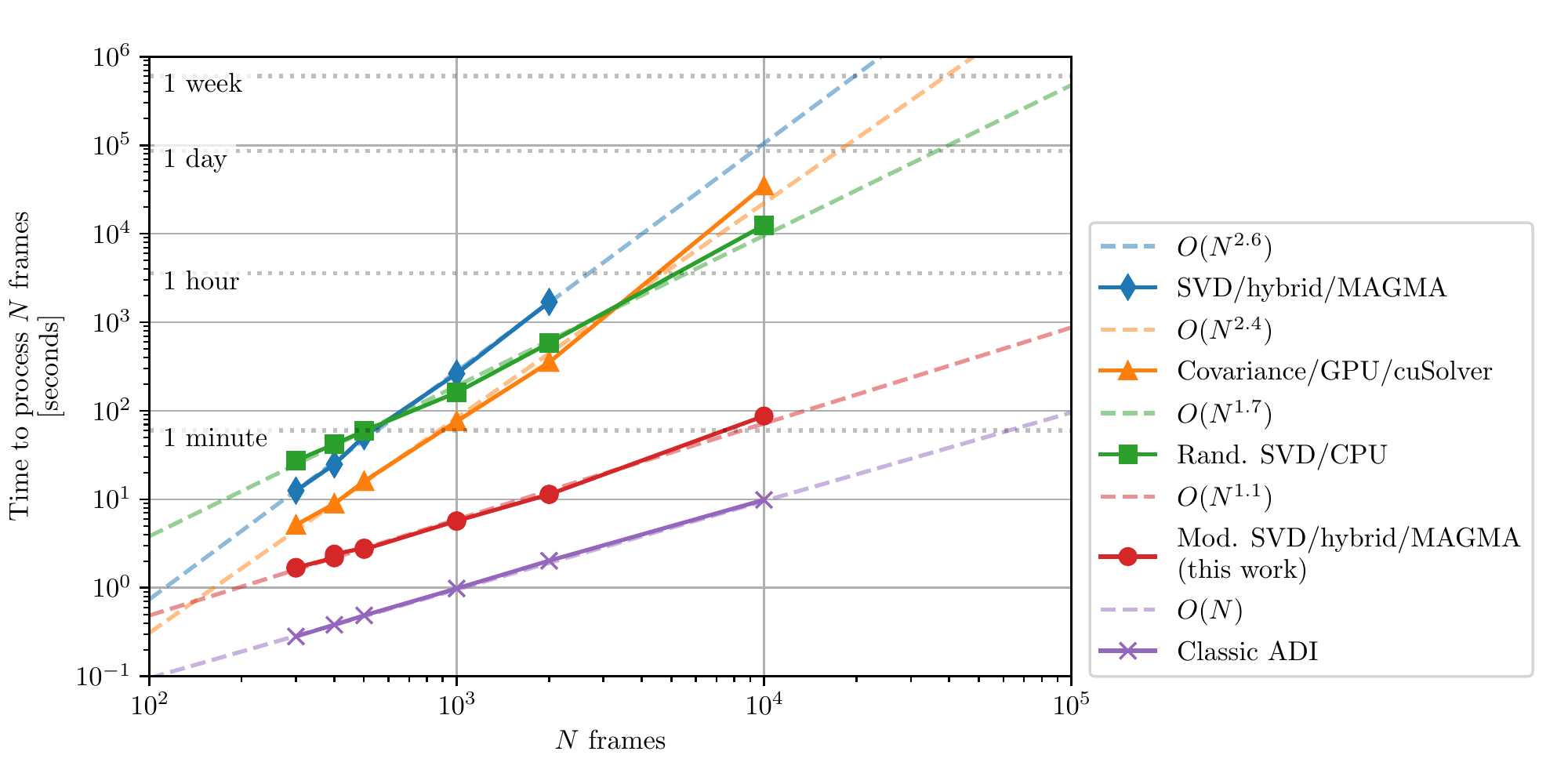}
    \caption{Actual runtimes for the algorithms tested. $N$ was varied by taking different slices from a 10,000 plane datacube. For algorithms that can take advantage of GPUs or alternative solvers, the combination that produced the best runtime at $N = 10^4$ frames was chosen to plot. After applying a mask to subset the data, the total number of pixels per plane used was $p = 6817$. Benchmarked on a UA HPC Ocelote cluster node with 250 GB RAM, an NVIDIA Tesla P100 GPU with 16 GB of GPU RAM, and a 28 core Intel Xeon E5-2680 v.4 CPU. \label{fig:benchmark}}
\end{figure*}

To validate this algorithm we developed a Python implementation of KLIP+ADI into which we could substitute different schemes for recomputing the SVD, as well as different solver routines.\footnote{Available as \cite{starbgone} and at\\ \href{https://github.com/joseph-long/svd-downdate-klip}{https://github.com/joseph-long/svd-downdate-klip}} After verifying that all algorithms produced equivalent signal-to-noise ratios on test data (Section~\ref{sec:compare_snr}), we compared execution times between KLIP implemented by:
\begin{itemize}
    \item finding eigenvectors of image-to-image covariance matrix $\mathbf{C}$ for each frame,
    \item computing the SVD of a new $\mathbf{R}$ for each frame,
    \item computing the randomized SVD of a new $\mathbf{R}$ for each frame following \cite{halko_finding_2011} using an implementation from {\tt scikit-learn} \citep{scikit-learn}
    \item or modifying the rank-$k$ SVD of $\mathbf{X}$ with our algorithm.
\end{itemize}

Where applicable, we compared between Intel MKL (CPU), MAGMA (CPU/GPU hybrid), and cuSolver (GPU) linear algebra solvers for the implementation of these algorithms. Timings at a variety of $N$ for $k = 200$ modes are shown in Figure~\ref{fig:benchmark}. In the interests of clarity, a subset of algorithm/solver combinations are shown (chosen to show each algorithm with the shortest runtime achieved, whichever device or library that may have used).

The characteristics of KLIP applied to differential imaging data, in which the references and target frames are drawn from the same set of observations, mean that our algorithm to perform $N$ updates to the SVD can outperform $N$ randomized SVD computations---which is the algorithm with the next-best asymptotic scaling exponent. The randomized SVD may nevertheless be useful for the initial computation of the rank-$k$ SVD that this algorithm modifies.

Classic rotate-and-stack ADI, in which a mean image is subtracted, the frames are rotated, and the images are combined by median or mean pixel values, scales linearly in $N$. Figure~\ref{fig:benchmark} shows that our SVD modification algorithm also exhibits (nearly) linear scaling over $300 < N < 10^4$.

Prior to developing this algorithm, we evaluated GPU computation as a possible path to speeding up KLIP. However, the general SVD solver for the GPU in cuSolver is notably slower than CPU and CPU/GPU hybrid algorithms. We note that cuSolver is slightly faster than MAGMA or MKL at solving symmetric eigenproblems at large $N$, but overall did not deliver a dramatic speedup to the existing algorithm. Fortunately, not only is CPU/GPU hybrid computation of the modified SVD algorithm faster, computers typically have far more RAM than GPU memory, allowing us to postpone investigating out-of-core algorithms.

Compared to the theoretical scaling for existing algorithms, execution time increases more slowly than expected with increasing $N$. This implies that our execution time is not dominated by number of floating point operations. Nevertheless, the predicted linear scaling is accurate to a good approximation for the modified SVD algorithm.

\section{Conclusions} \label{sec:conclusions}

We have described how to apply a mathematical identity to efficiently recompute the truncated SVD of a matrix after removing a subset its columns. We applied this to the computationally expensive problem of obtaining Karhunen-Lo\`eve eigenimages for starlight subtraction when the target and reference images are drawn from the same set of observations. This technique computes $\ZKL{}$ for each frame in an ADI sequence in $<1/2$ the time at $N = 300$ and $<1/140$ the time at $N = 10^4$ when compared with the most efficient algorithms. It exhibits near-linear scaling as $N \to \infty$, making it useful for large datasets. Our comparison of CPU and GPU implementations of the solver routines shows hybrid techniques are advantageous.

Computing modified KL bases for each frame of large-$N$ data cubes was formerly intractable due to the scaling properties of KLIP with varying reference sets. We believe this technique will enable analysis of large-$N$, short-integration data cubes that capture short-lifetime speckles without compromises. This in turn accelerates KLIP hyperparameter searches, such as for search region geometry and exclusion amount. Future work will investigate the starlight subtraction performance at large $N$ and $k$ for data cubes produced by MagAO-X and other ExAO systems.

\acknowledgments

JDL thanks Jennifer Lumbres for her assistance with debugging linear algebra notation on paper.

The authors thank the Heising-Simons Foundation (Grant \#2020-1824), NSF AST (\#1625441, MagAO-X), and NASA APRA (\#80NSSC19K0336) for their support. They would also like to thank the anonymous reviewer for helpful comments that improved the clarity of this work.

This work used High Performance Computing (HPC) resources supported by the University of Arizona TRIF, UITS, and the Office for Research, Innovation, and Impact (RII) and maintained by the UArizona Research Technologies department.

%

\vspace{5mm}


\software{NumPy \citep{numpy}, CuPy (\url{https://cupy.dev/}), MAGMA \citep{magma}, PyTorch \citep{pytorch}}, scikit-learn \citep{scikit-learn}, matplotlib \citep[][\citealt{matplotlib332}]{matplotlib}, IPython/Jupyter \citep[][\citealt{jupyter}]{ipython}, cuSolver (\url{https://developer.nvidia.com/cusolver}), Intel MKL (\url{https://software.intel.com/mkl})







\bibliography{references}{}
\bibliographystyle{aasjournal}



\end{document}